\documentstyle[12pt,aaspp4,flushrt]{article}
\def\capt{\small \baselineskip 12pt }

\def\eq#1{(\ref{eq:#1})}
\def\equ#1{equation~(\ref{eq:#1})}
\def\eqd#1{eq.~[\ref{eq:#1}]}
\def\se#1{\S\ref{sec:#1}}
\def\Figu#1{Figure~\ref{fig:#1}}

\def\cl{\centerline}
\def\\{\hfill\break}

\def\etal{{\it et al.\ }}
\def\rms{{\it rms\ }}

\def\eg{{e.g.}}
\def\ie{{i.e.}}

\def\vev#1{\langle#1\rangle}
\def\av{\vev}

\def\be{\begin{equation}}
\def\ee{\end{equation}}

\def\ifm#1{\relax\ifmmode#1\else$\mathsurround=0pt #1$\fi}
\def\kms{\ifmmode\,{\rm km}\,{\rm s}^{-1}\else km$\,$s$^{-1}$\fi}
\def\hmpc{\,\ifm{h^{-1}}{\rm Mpc}}
\def\hkpc{\,{\rm h^{-1}kpc}}

\def\dd{{\rm d}}

\def\msolar{M_{\odot}}
\def\hmsun{h^{-1}\msolar}

\def\ltsima{$\; \buildrel < \over \sim \;$}
\def\lsim{\lower.5ex\hbox{\ltsima}}
\def\gtsima{$\; \buildrel > \over \sim \;$}
\def\gsim{\lower.5ex\hbox{\gtsima}}

\def\pmb#1{\setbox0=\hbox{#1}%
 \kern-.025em\copy0\kern-\wd0
 \kern.05em\copy0\kern-\wd0
 \kern-.025em\raise.0433em\box0}

\def\v0{\pmb{$0$}}

\def\rs{R_{\rm s}}
\def\rhos{\rho_{\rm s}}
\def\rvir{R_{\rm vir}}
\def\mvir{M_{\rm vir}}
\def\dvir{\Delta_{\rm vir}}
\def\vvir{V_{\rm vir}}
\def\vm{V_{\rm max}}
\def\mstar{M_*}
\def\hfc{HFC }
\def\tm{\tilde M}
\def\tal{\tilde\alpha}
\def\tbe{\tilde\beta}
\def\tz{\tilde z}
\def\phib{\phi_{\rm b}}

\begin{document}

\medskip
\baselineskip 14pt

\title{Velocity and Mass Functions of Galactic Halos:\\
       Evolution and Environment Dependence} 

\author{Y. Sigad \altaffilmark{1},
T.S. Kolatt \altaffilmark{1},
J.S. Bullock \altaffilmark{2},
A.V. Kravtsov \altaffilmark{2,3},
A.A. Klypin \altaffilmark{4}, \\
J.R. Primack \altaffilmark{5},
\& A. Dekel \altaffilmark{1}
}

\altaffiltext{1}{Racah Institute of Physics, The  Hebrew University,
Jerusalem 91904, Israel}
\altaffiltext{2}{Department of Astronomy, Ohio State University,
Columbus, OH 43210}
\altaffiltext{3}{Hubble Fellow}
\altaffiltext{4}{Astronomy department, New Mexico State University,
Box 30001, Dept. 4500, Las Cruces, NM 88003}
\altaffiltext{5}{Physics Department, University of California, Santa
Cruz}


\begin{abstract}

We study the distribution functions of mass and circular velocity for dark
matter halos in N-body simulations of the $\Lambda$CDM cosmology, addressing
redshift and environmental dependence.  The dynamical range enables us
to resolve subhalos and distinguish them from ``distinct" halos. The mass
function is compared to analytic models, and is used to derive the
more observationally relevant circular velocity function.
The distribution functions in the velocity range 100--500$\kms$
are well fit by a power-law with two parameters, slope and amplitude.  We
present the parameter dependence on redshift and provide useful fitting
formulae.  The amplitudes of the mass functions decrease with $z$, but,
contrary to naive expectation, the comoving density of halos of a fixed
velocity $\sim 200\kms$ actually increases out to $z\sim 5$.  This is because
high-$z$ halos are denser, so a fixed velocity corresponds to a smaller mass.
The slope of the velocity function at $z=0$ is as steep as $\sim -4$, and
the mass and velocity functions of distinct halos steepen with increasing $z$,
while the functions of subhalos do not steepen with $z$, and become even
flatter at $z>2$. A simple observable prediction is that the slope of the
velocity function of isolated galaxies is steeper 
than that of galaxies in groups by as much as unity, reflecting the
density biasing of high-velocity 
halos. We confirm that the Press-Schechter approximation typically
overestimates 
the halo mass function by a factor of $\sim 2$, while modified approximations
provide improved predictions.

\end{abstract}

\subjectheadings{cosmology: theory --- dark matter --- 
galaxies: formation --- large-scale structure of universe}

\section{Introduction}
\label{sec:intro}

Understanding the evolution of structure in the universe and its relation 
to initial conditions is a fundamental issue in cosmology. 
Galaxy formation involves physical processes such as star formation,
supernovae feedback, and dust extinction which we still are far from
understanding in full. On the other hand, the clustering of the
collisionless dark-matter (DM) component is well understood.
DM particles cluster into halos, which subsequently accrete more mass 
either by gradual infall or by merging with other halos. 
An accurate description of the distribution of DM halo mass (the mass function)
and its temporal evolution are crucial for our understanding of the 
evolution of the luminous galaxies that reside in these halos.

A popular approximation to the mass function is provided by the
Press-Schechter formalism (Press \& Schechter 1974, hereafter PS). 
The initial density fluctuation field, smoothed on some comoving 
scale $R$, is assumed to evolve according to linear theory
until it reaches a critical value, $\delta_c=1.69$, at
which time it is assumed to have collapsed and virialized.
The critical value is determined by a linear extrapolation to the 
collapse time as set by the spherical collapse model.  Many of the 
simplifying assumptions behind the PS formalism do not seem to be 
physically motivated, yet it turns out to predict with reasonable accuracy the 
mass function as seen in simulations; the neglected effects tend to
cancel each other (Monaco 1998).

While being very useful for qualitative and semi-quantitative analyses, 
the accuracy of the PS approximation may not be enough when quantitative
details are concerned. For example, it has been found based on
simulations that for halo masses below $M=10^{13} - 10^{14}\hmsun$
(depending on the specific cosmological model) the PS approximation
overestimates the mass function by as much as a factor of $\sim 2$
(Gross \etal 1998; Lee \& Shandarin 1999; Somerville \etal 2000).
At the high-mass end, $M \gsim 10^{15} \hmsun$, the PS formalism 
underestimates the abundance of DM halos, especially at high redshift 
(\eg, Governato \etal 1999, Somerville \etal 2000).
Improved approximations have been proposed in order to correct these 
inaccuracies, \eg, a modification of PS by Sheth and Tormen 
(1999, hereafter ST), practically replacing the spherical collapse
model with an ellipsoidal collapse model (Sheth, Mo, \& Tormen 2000). 
Lee \& Shandarin (1998, hereafter LS) 
also use a non-spherical approach, based
on the Zeldovich approximation (Zeldovich 1970), to improve the PS
approximation. 

The need for a more accurate description of the mass function
is further highlighted by the development of a useful approach
for studying galaxy formation and evolution based on
semi-analytical models (Kauffmann \etal 1999; Somerville \& 
Primack 1999; Baugh \etal 1999).  Semi-analytical models try to
circumvent the complications associated with baryonic processes by  
simplified prescriptions and artificially embedding galaxies
within DM halos, which allows them to subsequently predict observable 
statistical quantities such as
the galaxy luminosity function or the Tully-Fisher relation.
Some of the important processes governing the 
evolution of galaxies in semi-analytical models are directly related
to the DM clustering properties; for example,
the mass function and merger histories of the halo populations are 
key ingredients in semi-analytical models (Kauffmann \etal 1999;
Somerville \& Kolatt 1999). 
These need to be followed with a better accuracy than provided by
the PS approximation. Furthermore, the PS formalism
does not address substructure within halos. When two halos
merge, the PS approximation immediately labels them as one virialized halo.
However, high-resolution simulations (e.g., Klypin
\etal 1999a, Ghigna et al. 1998) show that
substructure does maintain its identity after merging into larger
halos. This means that galactic halos are expected to survive tidal
stripping for some time, with possible important consequences for
collisions and starbursts (Kolatt \etal 1999, 2000).

An alternative to modeling the complicated physics needed to predict 
galaxy luminosities in individual halos is to calculate statistical 
properties such as the distribution function of the circular velocity 
of halos --- the velocity function.  By using observed luminosity-velocity 
relations, one can then relate the predicted velocity function to 
observational luminosity functions (Gonzales \etal 2000; Bullock \etal 2000b) 
and address halo number counts directly.
In addition, because modeling luminosities of the high-redshift galaxies
is even more uncertain than modeling the local population, 
the redshift evolution of the halo velocity function provides
a much-needed, and more direct handle on the galaxy population than
the corresponding mass function.

In order to obtain the velocity function we use a high-resolution
simulation which has the force resolution
and mass resolution necessary to determine the maximum circular
velocities of halos, and is capable of resolving substructure within
halos (Kravtsov, Klypin \&  Khokhlov 1997).
The simulation is complemented by a halo finding algorithm which also
classifies halos into nesting levels of halos within halos 
(Bullock \etal 2000a;
Bullock 1999;   see  \S\ref{sec:method}). Our halo finder/classifier
(HFC) fits each halo by an NFW density profile (Navarro, Frenk \&
White 1996) and automatically assigns a circular velocity profile to
each halo.
For the purpose of comparing to observations, some of the
halos are assumed to be galactic halos based on a simple 
prescription following basic observational constraints.

In \S\ref{sec:method} we present our method for extracting the
mass (\S\ref{sec:massf}) and velocity (\S\ref{sec:velf}) functions for the
different populations of halos (\eg, 
subhalos that reside in massive hosts, halos of galaxies
in groups, etc.), while correcting for incompleteness in the halo-finding 
algorithm by an iterative procedure. In \S\ref{sec:gal_id} we present
a simple scheme for identifying ``galactic'' halos.
In \S\ref{sec:results} we compare results for distinct halos 
to the PS, ST, and LS predictions, and extend our investigation to the
mass and  
velocity functions of substructure as well.  The results for halos and 
subhalos at $z=0$ are discussed in \S\ref{sec:z0}, and their redshift
evolution is presented in \S\ref{sec:evolution}. Results for
galactic halos in different environments are outlined in \S\ref{sec:gal_res}.
We discuss our results and compare them to other studies in \S\ref{sec:conc}.

\section{Method}
\label{sec:method}

Advances in numerical simulations of the collisionless DM component 
allow a dynamical range large enough for studying substructure within 
halos in a volume that approaches a fair cosmological sample of the
galaxy distribution (Klypin \etal 1999a). Using an adaptive refinement 
tree code (Kravtsov, Klypin \& Khokhlov 1997), we have simulated the
currently popular $\Lambda$CDM model within a comoving periodic box 
of 60$\hmpc$, with a force resolution of $f_{\rm res}\approx 2\hkpc$ in
the dense regions. The cosmological model is spatially flat,
with matter and cosmological constant contributions of
$\Omega_{\rm m} =0.3$ and $\Omega_{\rm \Lambda} =0.7$ at $z=0$. The Hubble
constant is $h=0.7$ ($H_0 \equiv 100h\kms\!\ {\rm Mpc}^{-1}$),
and the fluctuation amplitude today is normalized by $\sigma_{8} = 1.0$.
The $256^3$ simulated particles
imply a mass resolution of $m_{\rm p}=1.1\times 10^9\hmsun $.

Within this simulation it is possible to resolve halos inside halos
to four levels down the hierarchy. This is done via a halo finder/classifier,
based on the bound density maxima method (Klypin \& Holtzman 1997), 
which has been specifically developed for the purpose of analyzing 
substructure. A detailed description of the \hfc can be found in
Bullock (1999) and in the appendix of Bullock \etal (2000a). Below are 
some of its basic ingredients that are relevant to the mass and velocity 
distribution functions.

A key feature of the \hfc is that it models the radial density profile
of each halo with a universal functional form. We use here the NFW profile:
\be
\rho(r)={ {\rhos} \over {(r/\rs)(1+r/\rs)^2} } \ \   ,
\label{eq:nfw}
\ee
where the two free parameters are an inner density $\rhos$ and a 
characteristic scale length $\rs$. 
It is also convenient to define the virial radius of a halo by
\be
\rvir ={\left({{3\mvir }\over{4\pi\dvir \,\bar{\rho}}}\right) }^{1/3} \ \ ,
\label{eq:rvir}
\ee
where $\mvir$ is the virial mass, and virialization is defined by the 
mean density inside $\rvir$ being $\dvir$ times the mean density 
$\bar{\rho}$. The value of the virial overdensity $\dvir$ comes from the 
top-hat collapse model; it is about 200 for an Einstein-deSitter cosmology,
and $\dvir\simeq 340$ at $z$=0 for the $\Lambda$CDM model simulated here.
The corresponding virial velocity is defined by $\vvir^2 = G \mvir/\rvir$.
The profile fits were performed in the range $r > 0.02 \rvir$. The fitting
procedure automatically provides errors for the parameters $\rhos$ and $\rs$, 
which we then translate to errors in the other useful halo parameters such
as $\mvir$ and $\vvir$. 

The modeling of each halo by a smooth spherically-symmetric profile helps 
the identification of halos within halos. Halos at the top of the hierarchy
(not within other halos) are termed ``distinct'', while all other halos
are termed ``subhalos''. We only try to fit halos with modeled mass 
equivalent to at least 50 particles.
As will be seen below (\S\ref{sec:massf}), 
the \hfc finds halos with $M\gsim 2\times 10^{11}\hmsun$ at almost 100\%
efficiency, while at lower masses the constructed halo catalogs
gradually become incomplete.  
At $z$=0, a total of $\sim$8000 DM halos are identified, 90\% of which 
are distinct. Their numbers drop by approximately an order of magnitude 
for each level down the hierarchy.

In the following two subsections we present our method for
reconstructing the mass and velocity functions of any subset of the halos
identified by the HFC. An iterative procedure corrects for the
incompleteness at small masses (\S\ref{sec:massf}). The velocity
function is recovered in \S\ref{sec:velf} based on the 
derived mass function and the NFW profile fit.

\subsection{\bf Derivation of the Mass Function}
\label{sec:massf}

\begin{figure} [t!]
\vspace{9.0truecm}
{\includegraphics{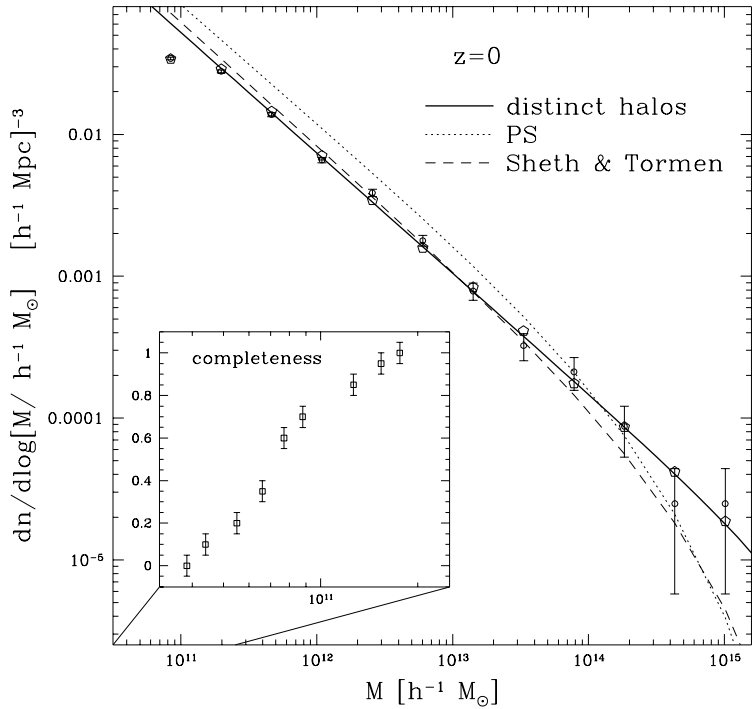}}
\caption{\capt
The mass function for distinct halos at $z$=0. 
The raw counts $N_j$ and their errors are shown by open circles with
error bars. 
The unbiased Schechter function is marked by the solid curve, and
the associated biased and incomplete counts $\tilde N_j$ are shown
as open pentagons; their proximity to the raw counts is a measure
of our success in recovering an unbiased mass function.
Also shown are the predictions of the Press-Schechter (dotted) and 
Sheth-Tormen (dashed) approximations.
Inset: The completeness function at the low-mass end.
}
\label{fig:massf_dist_comp}
\end{figure}

A simple count in bins of the halo population is prone
to two main sources of systematic error. One is due to the incomplete
efficiency of the \hfc at masses below $\sim 2\times 10^{11}\hmsun$,
which
causes an undercount in the low-mass bins.
The other is Malmquist bias due the low-mass cutoff imposed 
at 50 particles and the gradient in the mass function.
In order to obtain an unbiased mass function, we pursue an
iterative procedure as described in the following eight steps:

[1] The mass $M$ is the modeled virial halo mass $\mvir$.
We first impose a mass cutoff corresponding to 50 particles, \ie, 
$M_{\rm min} \simeq 5.5\times 10^{10}\,\hmsun$, 
and then count the halos in bins of constant
$\log M$ width. The logarithmic bin size, which is about 0.4 but
slightly varying from case to case,
is determined such that we have about 15 bins in the available mass
range.  We denote these raw counts by $N_j$. They are shown in
\Figu{massf_dist_comp} for the case of distinct halos at $z=0$.

[2] We evaluate the errors in each bin, $\Delta N_j$,
as the sum in quadrature of the
Poisson error due to the finite number of halos in the bin and the error in
the assignment of halos to bins because of the uncertainty in the
halo mass as determined by the HFC.
In order to evaluate the latter, we produce 20 synthetic halo catalogs
in which we perturb each mass of the simulated sample by an amount
drawn at random from a Gaussian distribution of width equal to the
corresponding \hfc error.
We count the halos in bins for each perturbed catalog,
and take the error corresponding to each bin to be
the standard deviation over the 20 perturbed catalogs.
The total errors, $\Delta N_j$, are shown as error bars in
\Figu{massf_dist_comp}. 

[3] We compute the completeness function of the HFC.  It is assumed to
depend only on the number of particles in the halo, namely $M$,
and not on redshift, or on whether the halo is a distinct halo or a
subhalo.
This assumption (which is validated a posteriori) allows us to
evaluate the completeness only once, \eg, for distinct halos at $z$=0,
and use it as is for all others cases.
We find that in an extended range below $10^{14}\hmsun$, the mass
function resembles a power law, while it drops sharply only
near the low-mass end where incompleteness effects are manifested.
We therefore assume that the underlying mass function is a power law
all the way down to $M_{\rm min}$ and determine this power law by
a fit that uses the bins and errors determined above but
excludes the first two bins near the minimum mass cutoff.
The completeness in the j-th bin, $C_j$, is defined by the ratio of
the raw count $N_j$ in that bin and the count corresponding to the power-law
fit.
The completeness function estimated this way is shown
in the inset of \Figu{massf_dist_comp}.

[4] The binned data corrected for incompleteness, $N_j/C_j$,
are fitted by a (temporary) Schechter function, according to which the
number density of halos in the mass range $(M,M+\dd M)$ is given by
\be
\dd n = \Phi (M)\,\dd M = \Phi_* \tm^{\alpha} e^{-\tm} \dd\tm
\ , \quad \tm \equiv M/M_* \ ,
\label{eq:schechter_m}
\ee
or
\be
\dd n = \phi (M)\,\dd (\log M) = \phi_* \tm^{\tal} e^{-\tm} \dd (\log M)
\ ,
\label{eq:schechter}
\ee
where $\tal=\alpha+1$ and $\phi_*=\Phi_* \ln 10$.
The fit parameters are $\tal$, $\phi_*$, and $M_*$, and the
minimum-$\chi^2$ fit procedure also provides the error in each one.
This completes the preliminary stage of the analysis.
Next, we correct the Schechter function for biases via an iterative
procedure.

[5]
Given an assumed ``true'' mass function in the $n$-th iteration $\phi_n$
(derived as above for the first iteration),
we produce 20 mock catalogs of halo masses drawn at random
from this Schechter distribution. We perturb the masses by a random deviant
that is drawn from a Gaussian distribution of errors appropriate for
that mass range, which we have derived before the
iterations from the actual error distribution within the corresponding
mass bin.  We then bin each perturbed mock dataset and fit it by a
Schechter function.
A ``biased" Schechter function $\phib$ is obtained by averaging the
Schechter parameters over the 20 noisy mock catalogs (and the errors in the
Schechter parameters are determined by the corresponding standard deviations
over the mock catalogs). Thus, the Malmquist biases lead from a ``true"
$\phi_n$ to an ``observed" $\phib$.

[6] To check how close we are to the desired solution, we pursue a
comparison of the current model and data in the ``observational" plane.
For this, we multiply
the biased Schechter counts by the completeness factors $C_j$, and
obtain biased and incomplete model counts in bins, $\tilde N_j$. The measure
used to evaluate convergence to the raw data is the weighted sum of residuals:
$S=\sum_j (\tilde N_j -N_j)^2 / (\Delta N_j)^2$. Once we succeed in
bringing $S$ to about unity, $\phi_n$ can be considered to be a good 
estimate of the unbiased mass function.

[7]
If $S$ is smaller than in the previous iteration, then
we correct our previous guess for the bias by:
\be
\phi_{n+1} = (\phi_n/\phib) \phi_n \ ,
\ee
and go back to step [5] for the next iteration.

[8]
When $S$ stops decreasing, we stop the iterations and adopt the current
$\phi_n$ as our unbiased mass function.
If $S$ is of order unity, this concludes our procedure.

We comment that the completeness function as estimated in step [3]
is in very good
agreement with the completeness function as estimated independently by
Bullock (1999), based on a comparison of our halo population with
the halos found by a straightforward halo-finding technique that
does not model the halo profiles.
This agreement provides strong evidence that the
estimate of the completeness function is correct.

The use of the Schechter function for the mass function is motivated by
the following two
arguments. First, luminosity functions are known to be well-fit
by the Schechter function, so if the mass-luminosity relation is roughly
a power law, then the mass function is expected to be of a similar
general form, with a power-law regime and a sharp drop at the large-mass
end.  Second, a similar function is predicted by the PS approximation
when the standard deviation of the density field $\sigma(M)$ is
assumed to be roughly a power law.
We will see that our data is capable of determining only the two
parameters of the power-law part, while $M_*$ is only weakly
constrained, and we don't even try to fine-tune the fit with any
additional parameters.

\Figu{massf_dist_comp} shows the final, unbiased Schechter function
(solid curve). Also shown are the corresponding biased and incomplete
counts, $\tilde N_j$ (open pentagons). 
The good agreement between these counts and the
raw counts $N_j$ is an indication for the success of our procedure
in obtaining an unbiased result. The final value of $S$ in this case
is 1.25, which is indeed on the order of unity. The PS approximation
overestimates the simulated mass function by a factor of
$\sim$2 for $M\lsim 5\times 10^{13}\hmsun$, then 
seems to cross over and
underestimates the mass function for $M > 10^{14}\hmsun$. The ST
approximation agrees with the simulated 
mass function to within 10\% for $M\lsim 5\times 10^{13}\hmsun$,
and seems to underestimate the simulated mass function for $M > 10^{14}\hmsun$.
The simulated mass function at $M>10^{14}\hmsun$ carries a large error
due to the small number of halos in this mass range, and therefore the
apparent discrepancy of the approximations in this range is of low
statistical significance.

\begin{figure}  [ht]
\vspace{9.0truecm}
{\includegraphics{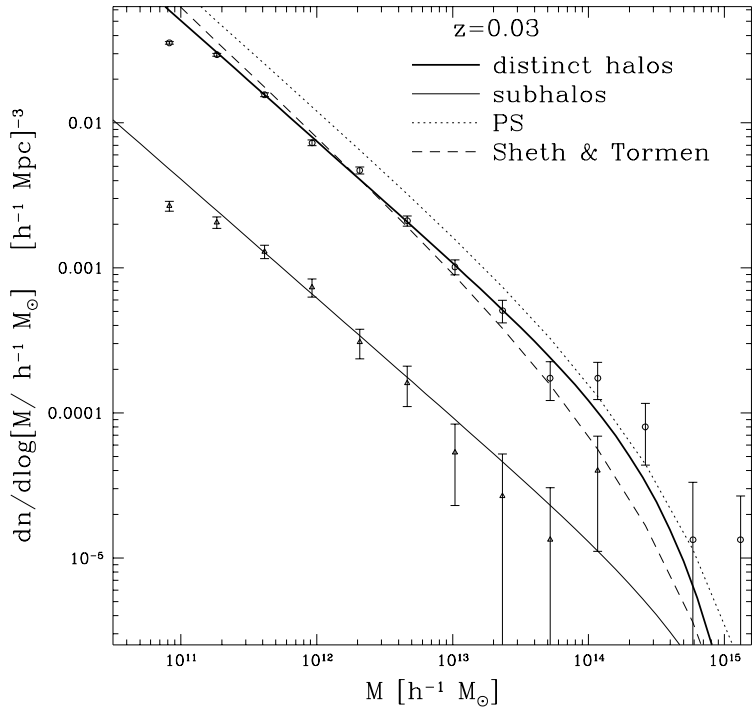}}
\caption{\capt
Mass function for distinct halos (top) and for subhalos (bottom)
at $z=0.03$. Raw counts are marked by symbols with error bars.
The curves are the unbiased Schechter-function fits.
The Press-Schechter (dotted) and Sheth-Tormen (dashed)
predictions for distinct halos are shown.
}
\label{fig:massf_dis_sub_003}
\end{figure}

\Figu{massf_dis_sub_003} compares the mass function for distinct halos 
and subhalos at $z=0.03$ (which is the same as $z=0$ for all practical
purposes).  The raw \hfc data are also shown, with the
apparent completeness turn-off at the two lowest mass bins.
The PS and ST approximations 
should be compared to the distinct halos' mass function.
The success of the approximations is very similar to the $z$=0 case.

The exponential turnoff set by $\mstar$ is not constrained very
well. In fact, the error in $\mstar$ is typically larger than $\mstar$
itself.  This is clearly demonstrated in the subhalo mass function of 
\Figu{massf_dis_sub_003}, where the last bin actually shows an apparent 
increase rather than a drop. The biased model counts are not shown in
this figure; their consistency with the raw counts is similar to what
we had in \Figu{massf_dist_comp}, with a similar value of $S$.

\subsection{\bf Derivation of the Velocity Function}
\label{sec:velf}

The \hfc also provides for each halo an NFW maximum circular velocity,
$\vm$, and one could straightforwardly compute a raw velocity function.
However, it would be impossible to correct this velocity function
for biases following a similar procedure to the one applied to the
mass function, because the incompleteness of the \hfc depends on mass,
and there is no one to one correspondence between velocity and mass. 
We therefore adopt a different approach, where the biases are corrected
at the mass-function level, and the velocity function is derived
from the corrected mass function.

The conditional distribution of $\vm$ given $M$ is not approximated
by a simple function and it strongly depends of redshift, so we 
choose to go from mass to velocity via the NFW ``concentration" parameter 
$c\equiv\rvir /\rs $.
We draw halo masses at random from the corrected Schechter mass distribution, 
and assign to each of them a value of $c$ drawn from the conditional 
lognormal distribution
\be
P(c|M)={{1}\over{\sqrt{2\pi}\Delta}}
\exp\left[ {{-(\log c - \av{\log c|M})^2}\over{2\Delta^2}} \right] 
\ \ ,
\label{eq:pcm}
\ee
where $\Delta$ is the conditional standard deviation of $\log c$ at a 
given $M$. The values of $\Delta$ and $\langle\log c|M\rangle$ 
are adopted from Bullock \etal (2000a); they are different at different
redshifts as well as for distinct halos and subhalos.
The lognormal distribution assumed in \eq{pcm} will be justified a
posteriori. 
Given $M$ and $c$, the NFW profile is uniquely determined, and in
particular
\be
\vm =0.465\,\vvir(M) \left[ c^{-1} \ln (1+c) - (1+c)^{-1} \right]^{-1/2}\ \  .
\label{eq:vmc}
\ee
We then count the halos in bins of constant $\log \vm$ width, and fit
a power law.

One main uncertainty in comparing the raw halo velocity function to
that of galaxies is the response of the halos to the dissipative contraction
of gas inside them. The typical effect is of an increase  
in halo maximum circular velocity, both due to the direct gravitational 
force exerted by the disk and the contraction of the halo in response
to the additional inwards pull by the formed disk.
We estimate the change in the relation between $M$ and $\vm$ 
using the fitting formula of Mo, Mao, \& White 
(1997, hereafter MMW), under the assumptions that \eq{pcm} is valid,
all halos have a spin parameter $\lambda=0.05$, the specific angular momentum
of the disk is equal to that of the dark matter, and the 
halo mass fraction which ends up in the disk is $m_d=0.04$. 
The counts of the corrected $\vm$ values are also well fit by a power law.

\begin{figure} [t!]
\vspace{9.0truecm}
{\includegraphics{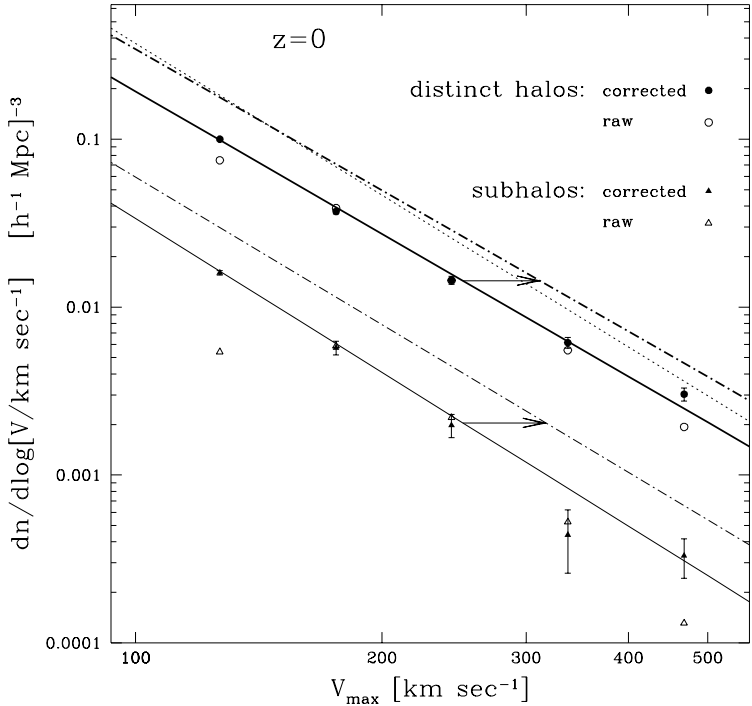}}
\caption{\capt
Velocity function for distinct halos (circles, thick lines) and
subhalos (triangles, thin lines) at $z$=0. Open symbols refer to the raw,
incomplete counts, while filled symbols with error bars
refer to the corrected counts derived from the mass function. The solid
lines are the power-law fits to the corrected counts, 
while the dot-dashed lines are the fits to the counts corrected
for baryonic infall. The dotted line is a prediction based on
the Press-Schechter formalism, not corrected for baryonic infall.
}
\label{fig:massf_dis_sub_vel}
\end{figure}

\Figu{massf_dis_sub_vel} shows the velocity functions
for distinct halos and for subhalos at $z$=0.
Shown are the incomplete, raw counts as well as the 
corrected counts as derived from the unbiased Schechter mass function.
The associated power-law fits  
are shown both before and after the correction for baryonic infall.
Note how well this procedure corrects for the incompleteness
at the low $\vm$ end. This result is still based on the same
completeness function derived in the previous section as a function of
mass, and it thus confirms its validity as well as the validity
of \eq{pcm}.

The velocity functions are clearly well fit by power laws across
the whole available velocity range.
We note, however, that a more realistic, semi-analytic treatment of 
disk masses $m_d$ may modify the power-law shape at the low and high 
velocity ends (Gonzalez \etal 2000).

Also shown in \Figu{massf_dis_sub_vel} is a simplistic,
PS-based theoretical prediction for the velocity function, not corrected
for baryonic infall. It refers to distinct halos,
assuming that the mass function is given by the PS prediction, and
that the mass and velocity are related via the same concentration
parameter for all halos, determined by the average $\av{\log c|M}$.
This approximation fails to reproduce the velocity function of the 
halos in our simulation by a factor of $\sim 2$ in amplitude and by
a wrong slope.

\subsection{\bf Identification of ``Galactic" Halos}
\label{sec:gal_id}

\begin{figure} [t!]
\vspace{9.0truecm}
{\includegraphics{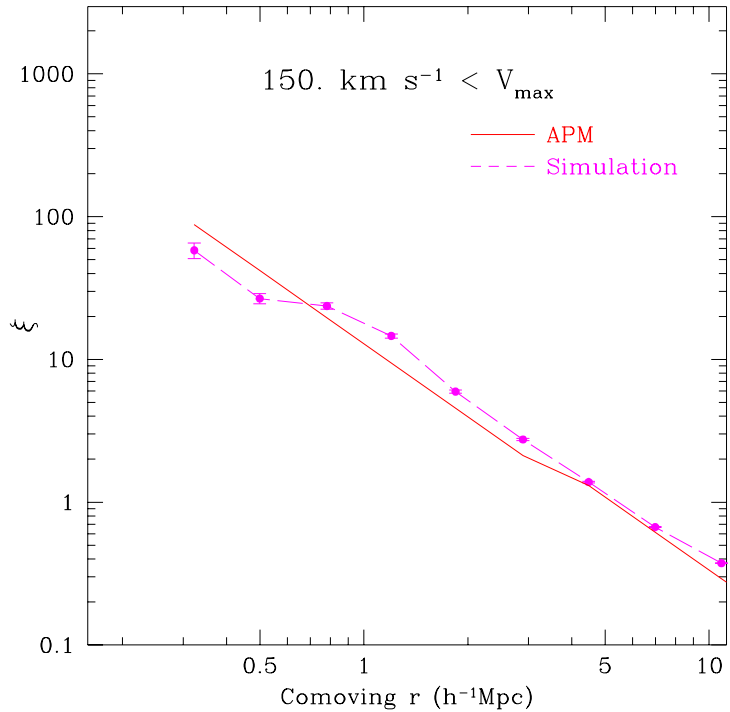}}
\caption{\capt
Two-point correlation functions for simulated ``galactic" halos with 
$\vm > 150\kms$ (symbols and dashed line) and for galaxies in the 
APM survey (solid line).
}
\label{fig:correlation_function}
\end{figure}

In order to directly compare our simulation results with 
data based on observed galaxy rotation curves, 
we should identify a fraction of our \hfc halos as ``galactic" halos.
We adopt the simplified assumption that
every halo with mass $M<10^{13}\hmsun$ hosts a galaxy. The mass cutoff
is placed to exclude groups and clusters from the galaxy count.
\Figu{correlation_function} shows the two-point correlation 
function of these ``galactic" halos; it is indeed in good agreement with  
the correlation function measured for galaxies in the APM survey 
(Baugh \& Efstathiou 1993) for $\vm > 150\kms$.

Further division into ``galaxies" in groups and clusters 
versus isolated, field ``galaxies" may yield interesting theoretical
predictions that can be directly confronted with observations.
We identify isolated galaxies with the distinct ``galactic" halos that 
do not host subhalos, and grouped galaxies with the ``galactic" halos 
that do contain subhalos or that are subhalos themselves.
Observationally, the grouped galaxies are 40 to 60\%
of the total galaxy population (\eg, Ramella, Pisani \& Geller 1997; 
Zabludoff \& Mulchaey 1998).
If we define a halo as a subhalo only if its center lies within the 
$\rvir$ of a larger halo (as done originally by Kolatt \etal 2000),
then the above classification scheme yields only $\sim 15\%$ 
in grouped ``galaxies" at $z=0$.  When we relax the proximity requirement 
of subhalos and hosts to $3\rvir$, we obtain in the simulations
a more reasonable fraction of $\sim 40\%$ in grouped ``galaxies".
Indeed, the observer classification tends to include loose groups whose 
DM halos may not overlap. 

For the two populations of ``galaxies'' we assign velocities according 
to the mass -- concentration relation of distinct halos. This may introduce
a small error in the case of grouped ``galaxies", which include some subhalos.

\section{Results}
\label{sec:results}

\subsection{\bf Distinct Halos and Subhalos at $z$=0}
\label{sec:z0}

\Figu{massf_dis_sub_003} shows the difference between the
mass functions of distinct halos and subhalos at the present 
epoch.\footnote{We actually use $z$=0.03 rather that $z$=0 in order to
compare neighboring redshift outputs (see \se{massf}) and to avoid
redundancy with \Figu{massf_dist_comp}.}
Apart from the difference in
normalization, the two populations seem to follow a similar power-law
distribution, with a slope of $\tal =-0.83\pm 0.03$ for the distinct
halos and $\tal =-0.82\pm 0.13$ for the subhalos. As mentioned
above, the turnoff scale $M_*$ is not well constrained, with a large 
formal error that is mainly due to the large Poisson errors in the high-mass
bins. The relative errors in the fit parameters grow only slightly with 
increasing $z$. For example, the relative
error in the slope increases by $\sim$20\% between $z$=0 and $z$=3.

We fit the velocity function with a power law:
\be
\dd n= \Psi(\vm)\,\dd \vm =\Psi_* \vm^\beta \,\dd \vm \ \ ,
\label{eq:vfitbeta}
\ee
or
\be
\dd n= \psi(\vm)\,\dd (\log \vm) =\psi_* \vm^{\tbe} \,\dd (\log \vm) \ \ ,
\label{eq:vfit}
\ee
with $\tbe =\beta + 1$ and $\psi_* =\Psi_* \ln 10$. We do not attempt
to fit a Schechter function because there is not even a hint for 
a break at large velocities; the weak signature of an exponential break 
seen in the mass function is completely erased in the velocity function by
the scatter in the concentration parameter (\eqd{pcm}).

Figure~\ref{fig:massf_dis_sub_vel} shows that the main
qualitative results of the mass function carry over to the velocity
function, namely, an order-of-magnitude difference in the
normalization but very similar slopes for distinct halos 
($\tbe=-2.82\pm 0.03$) and subhalos ($\tbe=-3.04\pm 0.10$).
This is despite the fact that the corresponding distributions
of concentration parameter [$P(c|M)$] are different.
The fits are quite robust, with the relative errors in the slope
and the normalization (defined at $V_{\rm max}=300\kms$)
limited to a few percent, at all redshifts in the range 0-3.

The effect of applying the simple MMW
baryonic infall prescription with $m_{\rm d} =0.04$ is a logarithmic
shift of the halo population to larger velocities 
(as indicated by the arrows). The corresponding increase 
in normalization is a factor of $\sim2$, both for
distinct halos and subhalos. The change in slope, to
$\tbe=-2.79\pm   0.04$ for distinct halos and
$\tbe=-2.92 \pm  0.09$ for subhalos, is statistically insignificant.
This was not expected {\it a priori} because the baryonic-infall correction 
depends on halo concentration $c$ (halos of higher $c$ are assumed to 
accrete more baryonic mass) and it is therefore expected to vary 
as a function of mass and velocity.
Furthermore, 
the correction is virtually the same for distinct halos and subhalos,
despite the fact that the relation between mass and concentration is
somewhat different for these two populations (Bullock \etal 2000a).
A more sophisticated treatment of infall,
which would take into account effects such as the lower fraction of baryons 
involved in forming the disk inside low-$V_{\rm max}$ halos, 
may yield a change in shape (e.g., a slight turn-off at low velocities, 
see Gonzalez \etal 2000).

\subsection{\bf Redshift Evolution of Distinct Halos and Subhalos}
\label{sec:evolution}

\begin{figure} [t!]
\vspace{7.00truecm}
{\includegraphics{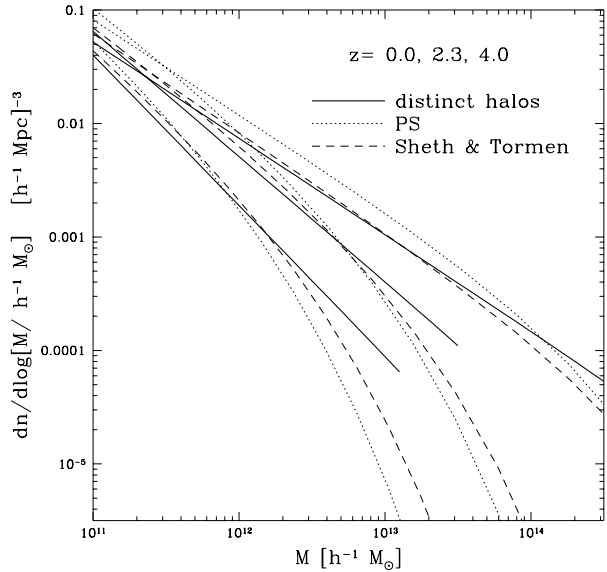}}
\caption{\capt
The fitted mass function for distinct halos in the simulations (solid)
at three different redshifts ($z$=0, 2.3, 4, from top to bottom). 
The fits are truncated to span a mass range defined by bins 
for which $N_j / \Delta N_j > 2$.
The approximations of PS (dotted) and ST (dashed) are shown for
comparison.
}
\label{fig:massf_3z}
\end{figure}

The mass functions for distinct halos at three different redshifts
in the range 0 to 4 are shown in \Figu{massf_3z}. Shown for comparison 
are the predictions of the PS and ST models. At $z$=0, we recover the 
already known result (\eg, Gross 1997, Somerville \etal 2000)
that at low masses the PS approximation overestimates the mass function 
by a factor of $\sim$2, while at high masses it is an underestimate. 
The ST approximation fares better 
in general, and especially at low masses; it is accurate to $<10\%$ up 
to $M\approx 5\times 10^{13}\hmsun$
at $z=0$, and $M\approx  2\times 10^{12}\hmsun$ at $z=4$.  
The crossover mass, where PS agrees with  the simulated mass function, shifts 
from $\sim10^{14}\hmsun$ at $z$=0 to $\sim 10^{12}\hmsun$ at $z$=4.  
By comparison, the typical collapsing mass ${\mathcal{M}} _\star$, defined by
$\sigma[{\mathcal{M}}_\star(z)]=1.69$ 
(where $\sigma[M]$ is the linear \rms
density fluctuation on a scale corresponding to mass $M$), is equal to 
$\sim 10^{13}\hmsun$ at $z$=0 and to $\sim 10^{9}\hmsun$ at $z$=4.

\begin{figure} [t!]
\vspace{9.0truecm}
{\includegraphics{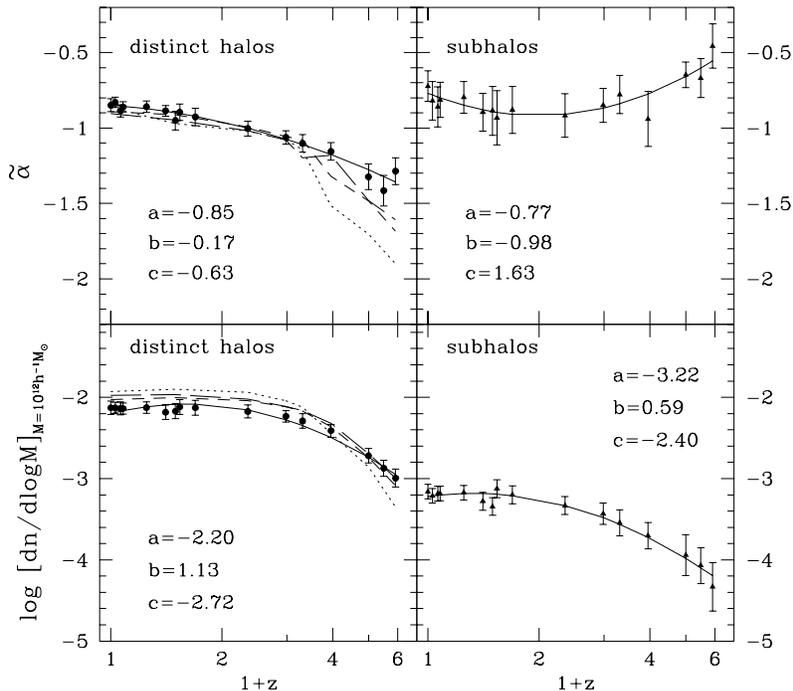}}
\caption{\capt
Redshift evolution of the mass functions of distinct halos and
subhalos. The top panels show the evolution in the slope of the
mass function (left -- distinct halos, circles; right -- subhalos,
triangles), and the bottom panels show the overall normalization at
$M=10^{12}\hmsun$. Functional fits to the redshift evolution are shown
as solid lines, where $a$, $b$,
and $c$ are the fit parameters (see text).
Also shown are the Press--Schechter (dotted), Sheth--Tormen
(dashed) and Lee--Shandarin (long-dashed)
predictions (relevant only for distinct halos).
}
\label{fig:massf_redshift}
\end{figure}

The redshift evolution of the mass functions for both distinct
halos and subhalos are displayed in \Figu{massf_redshift}.
We show the evolution of the Schechter fit parameters $\tal$ and $\phi(M)$
at $M=10^{12}\hmsun$.
We present the normalization by the latter rather than by
$\phi_*$ because the large scatter in $\mstar$ induces a non-negligible 
scatter in $\phi_*$ ($\mstar$ is determined by the value of $\phi$ at
$\mstar$),  
and because at this relatively low mass we have good statistics
throughout the studied redshift range.
The values of $\mstar$ and $\phi_*$ are given in Table 1.

The top panels of Figure~\ref{fig:massf_redshift} show the evolution in
the slope $\tal$ of the mass function of distinct halos and
subhalos. A marked difference is seen between the two kinds of halos.
While the distinct-halo mass function steepens with
redshift from $\tal=-0.85\pm 0.04$ at $z$=0 to $\tal= -1.3\pm 0.1$ at $z=5$,
the subhalo slope 
has a broad minimum of $\tal=-0.9\pm 0.15$ at $z\sim 1$
and it then rises to $\tal$=$-0.5\pm 0.15$ at $z$=5. 
A useful functional fit that traces this evolution is provided by the
quadratic polynomial 
\be
\tal(z)=a+b\tz+c\tz^2\ , \quad  
\quad \tz\equiv\log(1+z) \ . 
\label{eq:poly}
\ee
these fits are shown in the figure, and the corresponding parameters
$a$, $b$, and $c$ are listed in each panel.

The evolution of the normalization of $\phi$ at $10^{12}\hmpc$
is shown in the bottom panels of \Figu{massf_redshift}. 
Despite the big difference in amplitude, the trends with redshift
are similar for distinct halos and subhalos,
both remaining roughly constant between $z=0$ and $\sim 1$,
and then gradually declining, by roughly
an order of magnitude at $z$=5.
Indeed, the functional fits to the two kinds of halos show
a difference of $\sim 1$ in the offset $a$ and only 15\%
in the curvature $c$.
The error in the normalization at
$10^{12}\hmpc$ is much smaller than the error in $\phi_*$. 
The error is the sum in quadrature of the random error associated with
the count in the bin containing $M=10^{12}\hmpc$ and the systematic error
estimated by the change in the value of the fitted function across the bin.

For distinct halos we show in comparison the evolution of the
PS model predictions, fitted at each redshift by a Schechter mass function 
in the range  $10^{11}\hmsun<M<10^{15}\hmsun$.
The PS approximation overestimates the normalization at
$10^{12}\hmsun$ for $z<2$ by a factor of $\approx 1.7$, while at $z$=5
it is an underestimate by a factor of 
$\approx 2$. 
The slope $\tal$ is 
well approximated up to $z\sim2.5$, but is underestimated at higher
$z$ by as much as unity.  The ST and LS 
approximations fare better than PS. They are quite similar, predicting 
accurately the slope up to $z\approx 3$. At $3<z<5$ both
approximations predict a slope steeper by $\sim$20\% than what we find in
the simulations. The ST and LS models also improve on PS
by predicting the normalization (at $10^{12}\hmsun$) more accurately,
to within 25\% in the range $0<z<3$. The discrepancies in the
predicted slope (at high $z$)
and normalization (at low $z$) are small relative to the PS model, but
are, nevertheless, statistically significant.

\begin{figure} [t!]
\vspace{9.0truecm}
{\includegraphics{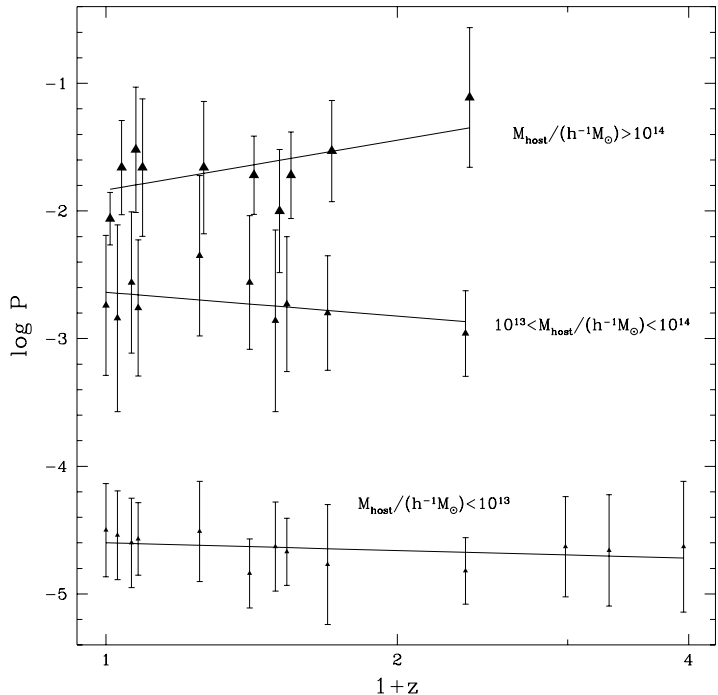}}
\caption{\capt
The conditional mass function: conditional probability density
$P(M_{\rm sub}|M_{\rm host})$ at subhalo mass of
$M_{\rm sub}=10^{12}\hmsun$. 
The host masses are divided into three mass bins:
$M_{\rm host} < 10^{13}\hmsun$ (small triangles)
$10^{13}\hmsun < M_{\rm host} < 10^{14}\hmsun$ (medium triangles),
and $M_{\rm host} > 10^{14}\hmsun$ (big triangles). The solid lines
are linear fits to the data.
}
\label{fig:massf_cond}
\end{figure}

We are also able to investigate the properties of a conditional
mass function,\footnote{Not to be confused with the
temporal conditional mass function, as in
the extended PS formalism, which is the probability of a
progenitor of mass $M_1$ at $z_1$ given a halo of mass $M_2$ at
$z_2$ (\eg, Lacey \& Cole 1993).}
defined as the probability density $P(M_{\rm sub}|M_{\rm host})$
of finding a subhalo of mass $M_{\rm sub}$ inside a host halo
of mass $M_{\rm host}$. 
The conditional mass function was derived along the lines of the
prescription described in \S\ref{sec:massf}.
The conditional mass functions were fit 
to a Schechter function at each $z$ output, and
the fit parameters are given in Table 2.
We have divided the host masses into three mass ranges bordered by
$10^{13}$ and $10^{14}\hmsun$, and compared the conditional mass
functions in these three ranges.
The slopes of these functions are found to fluctuate in the range 
$-0.9<\tal <-0.6$, with no obvious correlation with the host mass. 
There is also no evidence for redshift evolution of the slopes
(Table 2, second column), but note the large Poisson errors.

\Figu{massf_cond} shows the conditional probability density
$P(M_{\rm sub}|M_{\rm host})$ at $M_{\rm sub}=10^{12}\hmsun$, 
for the three ranges of $M_{\rm host}$, as a function of redshift.
The
normalization here is by the number of host halos (rather than by volume),
so that an integration over all subhalo masses per given host-halo mass
yields a probability of unity.
The linear fits to the data exhibit very little evolution, in concordance 
with the results for the whole subhalo population. 
We see no evidence for $M_{\rm host}$ dependence in the evolution of 
the subhalo mass function.  This could be useful for
constructing synthetic subhalo populations for semi-analytical simulations.
Since host mass correlates strongly with the subhalo environment density 
(e.g., top-hat smoothed at 1.5$\hmpc$, Bullock \etal 2000a), \Figu{massf_cond} 
also reflects the environment dependence of the subhalo mass function.

\begin{figure} [t!]
\vspace{9.0truecm}
{\includegraphics{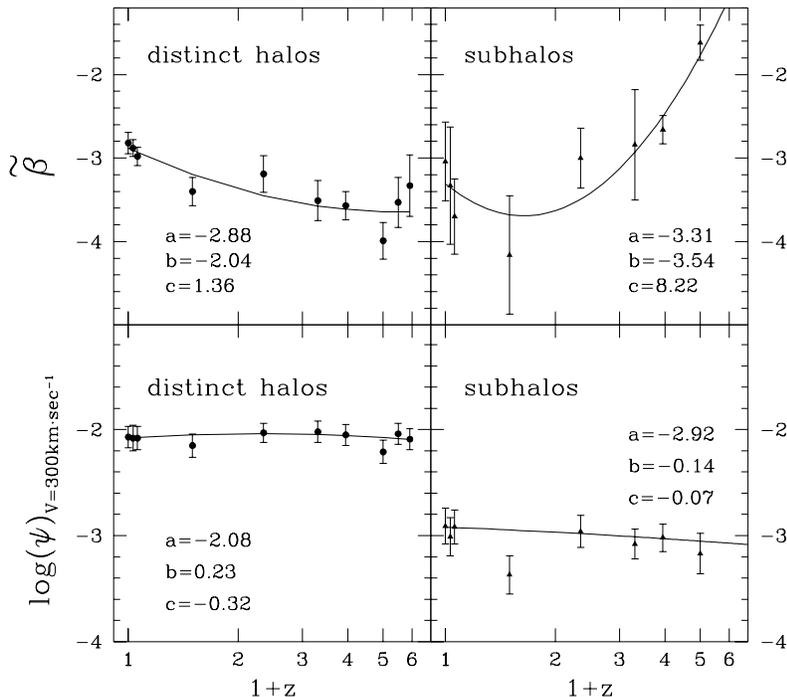}}
\caption{\capt
The redshift evolution of the velocity function for distinct halos
(left panels) and subhalos (right panels):
The top two panels show the evolution in the slope of the
velocity function, and the bottom panels show the overall normalization
at  $V_{\rm max}=300\kms$.  Functional fits to the redshift evolution
are shown as solid lines, and $a$, $b$,
and $c$ are the fit parameters (see text).
}
\label{fig:velf_redshift}
\end{figure}

\Figu{velf_redshift} shows the evolution of the parameters characterizing 
the velocity functions for distinct halos and subhalos. 
Here, as for the mass function, the normalization is characterized by
the value of the velocity function at a fixed velocity, of $300\kms$,
rather than by $\psi_*$.
The fit parameters $\tbe$ and $\psi_*$ of \equ{vfit} are listed in Table~3.
We also provide fits to the evolution by the same quadratic polynomial as 
for the mass functions, \equ{poly}.
The errors are estimated as for the mass function; they are larger
for subhalos because there are fewer of them.

The evolution of the slopes of the velocity functions is qualitatively
similar to that of the mass function. 
This is not surprising since the $M-\vm$ relation at all redshifts is
close to a power law, $M \propto \vm^{s}$, implying
$\tbe(z) \simeq s(z) \tal(z)$.  For distinct halos, 
$s=3.4$ at low $z$ and it approaches $s=3$ at high
$z$ as typical concentration values fall.  For subhalos, $s=3.9$ at
$z=0$ and it again approaches $s=3$ at high $z$ (Bullock et al. 2000a).

On the other hand, the normalizations of the velocity functions remain
roughly constant with redshift, unlike the decrease shown by the mass
functions between $z=1$ and $5$.
This interesting behavior is a combination of two effects.
First, ignoring the redshift dependence of $\dvir$, the
typical halo density increases with redshift roughly as $(1+z)^3$,
implying that halos of a given velocity correspond to objects of
smaller masses at high $z$:   
$M(\vm={\rm const}) \propto (1+z)^{-3/2}$.    
The increasing halo density with redshift is counteracted somewhat by
the tendency of high-$z$ halos to be less concentrated 
(Bullock \etal 2000a).  The net result is a very weak evolution in the 
comoving number density of halos of a given, high velocity.
Since the slope of the velocity function becomes steeper at high
$z$, the fact that the normalization at a fixed $\vm=300\kms$
is roughly independent of redshift implies that low velocity 
($\vm \lsim200\kms$) halos, and the galaxies that reside within them,
are predicted to be more abundant at high redshift.

\begin{figure} [b!]
\vspace{8.0truecm}
{\includegraphics{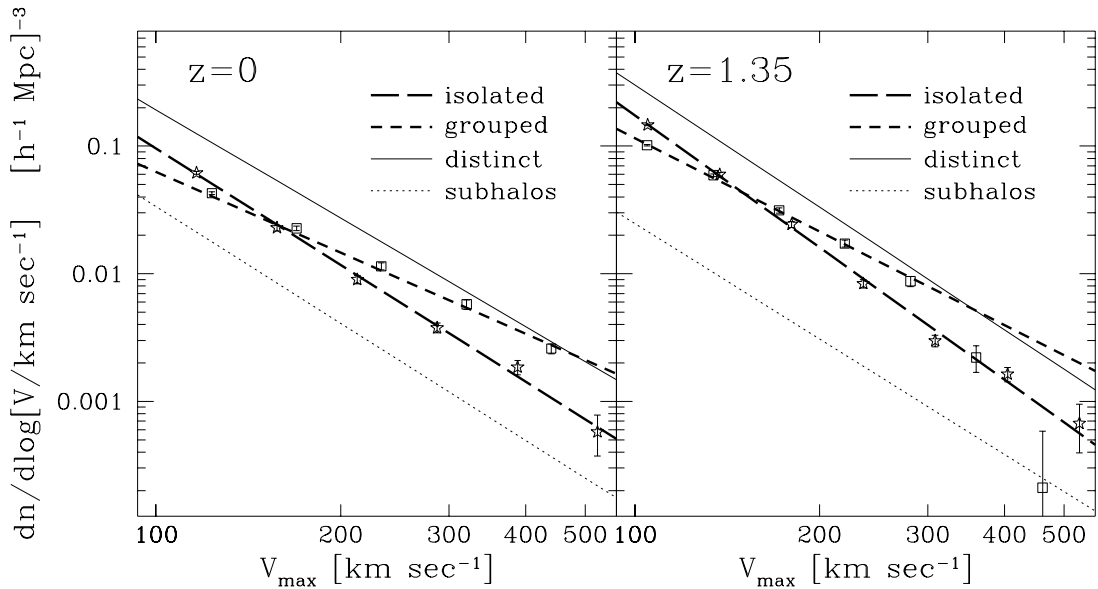}}
\caption{\capt
The velocity function for isolated galaxies (squares, short-dashed fits)
and grouped galaxies (stars, long-dashed fits) 
at $z$=0 and 1.35. also shown for reference are the velocity functions 
of distinct halos and subhalos (solid and dotted lines).
}
\label{fig:massf_galaxies}
\end{figure}

\subsection{\bf Isolated versus Grouped Galaxies}
\label{sec:gal_res}

In this subsection we aim at a prediction that can be directly 
tested observationally.
As described in \se{gal_id}, we crudely assume that each halo 
of mass smaller than $10^{13}\hmsun$ has a luminous galaxy in it;
this includes both distinct halos and subhalos.
We then divide the galaxy population into two classes, of
{\it isolated} galaxies and {\it grouped} galaxies, containing at $z=0$ 
about $60\%$ and $40\%$ of the galaxy population respectively.

\Figu{massf_galaxies} compares the velocity functions of these two
classes, at $z=0$ and at $z=1.35$.
The two functions display significantly different slopes,
with $\tbe=-3.04\pm 0.10$ and $-2.11\pm 0.03$ for isolated and
grouped galaxies at $z=0$, and with $\tbe=-3.44\pm 0.08$ and
$-2.43\pm 0.05$ at $z=1.35$.
The ratio between the two slopes at these two redshifts is virtually the
same, $\approx$0.7. At $z$=3 (not shown), the difference between the
slopes of the velocity functions is even larger:
$\tbe=-3.96\pm 0.05$ and $-1.85\pm 0.05$ respectively, a ratio $\approx$0.45.
It is worth mentioning that
the fraction of grouped halos (according to the $3 R_{\rm vir}$ criterion 
of \se{gal_id}) decreases from 40\% at $z=0$ and $1.35$ to 30\% at $z=3$.

In general, the velocity functions are well fit by power laws.
The exception is the velocity function of grouped galaxies at $z$=1.35,
which drops more steeply at $V_{\rm max} > 300\kms$.
This is an artifact of our imposed strict upper bound on galactic-halo
masses at $10^{13}\hmsun$, which translates into a smoother drop in 
velocity due to the scatter in the mass-velocity relation (\S\ref{sec:velf}).
The corresponding drop at $z=0$ is pushed to higher velocities.
In the real universe we may expect galactic halos more massive than 
$10^{13}\hmsun$, and therefore the drop may be unphysical and
should be ignored in the power-law fit.

The predicted difference between the slopes of the velocity functions
of isolated and grouped galactic halos reflects the ``biasing" tendency for
large halos to be more clustered; high-mass, high-velocity halos are
more likely to have companions and to exist in groups than their
low-mass, low-velocity counterparts. This relation carries over to
luminous galaxies, if we assume that every halo that obeys our simple 
mass criterion hosts a galaxy. The predicted slope difference  
between groups and the field should be observable, both at low and 
high redshifts.

\section{Discussion and Conclusions}
\label{sec:conc}

We have studied the distributions of mass and maximum circular velocity
of DM halos in a cosmological simulation of the $\Lambda$CDM model.
The mass function is useful for semi-analytic modeling of galaxy formation,
while the velocity function can be confronted with observations once
a relation between halo velocity and disk rotation velocity is assumed.
The high resolution allows us to address distinct halos as well as 
subhalos, at a range of redshifts, and to distinguish between field 
and grouped galactic halos.

We find that the Schechter function, and in particular its power-law
behavior over a wide range below the characteristic mass $M_*$, 
is a good fit to the different halo distributions of mass and velocity, 
over the whole range studied in this simulation, and at all times monitored.
This is once the incompleteness of the halo finder at small halos 
and the associated errors are properly accounted for.
Each of these distribution functions is therefore characterized by
a slope and an amplitude. The characteristic large scale where the 
function bends is not constrained properly because we do not sample 
enough large halos.  At the low end, we do not attempt to recover 
the velocity function below $100\kms$, and therefore cannot address 
possible deviations from a power-law at low velocities (Klypin \etal 1999b;
Moore \etal 1999).
At $z=0$, for distinct halos, we find mass and velocity function
slopes of $\alpha=-1.85\pm0.4$ 
and $\beta=-3.82\pm0.03$ in \equ{schechter_m} and \equ{vfitbeta} respectively.

We measure the time evolution of the mass and velocity
functions via the redshift dependence of the slope and amplitude
parameters.  The $z$ dependences are described for convenience by
simple functional fits out to $z=5$ 
[Figures~\ref{fig:massf_redshift} and \ref{fig:velf_redshift}], 
to be used in semi-analytic models of galaxy formation.  
The slope of the velocity function for distinct halos steepens with
redshift like 
$\beta=-3.9 -2.0\tz +1.4\tz^2 \pm0.2$, where $\tz\equiv \log(1+z)$. 
A particular prediction that should be tested observationally is that
the amplitude of the velocity function at fixed $\vm =300\kms$
hardly varies with redshift, for all types of halos.  This implies
that the number density of halos with $\vm \sim 200\kms$ and below is
actually predicted to {\it increase} slightly with increasing
redshift, as opposed to the naive expectation based on the way the
mass function evolves.  
 
This prediction for the weak 
evolution of the velocity function can also
be tested indirectly.  For example, by combining this result with the
observed evolution of the luminosity function of spiral galaxies, one
can deduce the way the Tully-Fisher relation might evolve with redshift 
(cf. Gonzalez et al. 2000; Bullock \etal 2000b).
To make this connection between the luminosity function and
the velocity function using the Tully-Fisher relation,
it will also be necessary to take into account the effects
of baryonic infall.  But this is important, since an inconsistency
between this prediction and direct observations of Tully-Fisher
evolution may question the validity of the $\Lambda$CDM model.

Most earlier cosmological simulations, and analytic approximations such as
Press - Schechter, overlook substructure within halos, which is 
bound to play an important role in galaxy formation.
Our simulations and halo finder enable a study of the hierarchical
halo population within a cosmological volume.
We find that the subhalo distributions can also be fitted by power laws,
though the errors are larger because there is an order of magnitude fewer
subhalos than distinct halos.
At redshifts of order 2 and beyond, we find that the subhalo population 
evolves differently from the distinct halo population.
While the slope of the distinct-halo functions becomes steeper 
with increasing redshift, the slope of the subhalo becomes flatter 
at higher redshifts.
On the other hand, the amplitudes of the subhalo and distinct-halo
functions decrease with increasing redshift in a similar way.
 
Several other simulations were used to study the evolution of substructure
within individual clusters of galaxies (\eg, Ghigna \etal 1998, van den Bosch
\etal 1999, Sensui \etal 2000, Okamoto \& Habe 1999).
The results of Okamoto \& Habe (1999), in particular, are quite similar 
to what we find. They see very little evolution of the subhalo mass function 
in the range $0<z<2$, with a power-law slope $\tal =\alpha+1 =-0.6$.
This similarity is despite the fact that they concentrate on a 
constrained realization of a 3-$\sigma$ density peak within an $\Omega=1$
SCDM model and their mass range is quite different.

For the purpose of semi-analytic modeling of galaxy formation, we also 
considered the substructure as a function of host-halo mass.
We find that except for the natural ``normalization" trend
(that more massive halos harbor a larger number of subhalos,
see the study of the multiplicity function in Kolatt \etal 2000),
there is only little evolution in the subhalo mass function
for any given host-halo masses.
These conditional mass functions can serve to improve the
semi-analytical recipes for placing galaxies in simulated DM halos.

Our derived velocity function for distinct halos agrees with the preliminary
independent analysis of the same simulation by Gottl\"{o}ber \etal (1998).
We improve on the preliminary analysis by correcting for small-mass 
incompleteness, placing meaningful error bars on the fit parameters, 
following the velocity function with higher time resolution and for a longer
history (up to $z$=5), and, in particular, extending our investigation 
to subhalos, and studying environmental dependence.
The derived velocity function for halos identified as galactic halos,
which is similar to that of distinct halos, is also in good agreement 
with Klypin \etal (1999b), who identified galaxies in the same
simulation, as well as in a simulation of higher resolution, 
complete down to $\approx 30\kms$. They used a different halo finder 
and assigned circular velocities in a different way. 
They report a velocity-function slope
which corresponds to $\tbe=-2.75$ and $\log \psi_*\approx 4.84$, 
in good agreement with our values for distinct halos. 
This agreement is encouraging evidence for the robustness of our results.

One of our most interesting predictions 
to be tested observationally
is that galactic halos in groups should have a significantly 
flatter velocity function than more isolated galactic halos.
The implication is that galactic halos at the high velocity end
should show a stronger tendency to reside in denser environments.
A detailed comparison to observations must incorporate the relation 
between the halo velocity and that of the luminous galaxy, which we 
crudely assumed here to be identical.
White, Tully \& Davis (1988) and Mo \& Lahav (1993) already found 
hints for a correlation between galaxy velocity and local galaxy density,
which seem to be qualitatively consistent with our predicted trend.
However, a proper quantitative comparison is yet to be done, 
treating effects of incompleteness as a function of velocity,
and considering in particular the {\it slope} of the velocity function
as a function of the environment.
This comparison should be done with larger, more complete datasets,
both at low and high redshifts.

High resolution N-body simulations accompanied by simple schemes for
galaxy identification  
can thus provide a powerful tool for investigating 
galaxy formation, especially those aspects of the problem that are not yet
properly addressed by semi-analytic modeling or by full hydro simulations.
Future simulations of larger volumes, which will include more massive
halos of cluster size, will permit better constraints on $M_*$.
Better statistics will also mean a more accurate
evaluation of the subhalo mass and velocity functions, especially
the conditional mass function, which could give us a better
handle on the relation between host halos and their internal substructure.

As a by product, our results can serve to evaluate analytic
approximations of the mass function in the clustering process.
At low redshift, we confirm earlier findings that the Press-Schechter 
approximation overestimates the true mass function by a factor of two 
at low masses, and underestimates it at high masses. This is in good agreement
with Gross \etal (1998) and Gross (1997), and in qualitative agreement
with Somerville \etal (2000) and Lee \& Shandarin (1999). This
agreement between the simulations, using a variety of halo-finding algorithms,
provides an additional confirmation for the robustness
of our \hfc in finding distinct halos.
The latter differ somewhat in the value of the mass where the PS
approximation coincides with the true mass function, probably due to
the different cosmologies they investigated ($\Omega =1$ SCDM and $\tau$CDM).
We find that the approximations of PS as well as ST and LS 
overestimate the normalization of the mass function at $z<3$ and
underestimate the slope at $z>3$, but the discrepancies between the
simulation and the ST and LS predictions are much smaller than 
the deviations of the PS formalism. All three approximations predict a higher 
rate of evolution for the mass function than seen in the simulations.

\section*{Acknowledgments}
The simulations were performed at NRL and NCSA. This work was
supported by grants from the Israel Science Foundation at HU, the US-Israel 
Binational Science Foundation at HU and UCSC, 
and NASA and NSF at UCSC and NMSU.
JSB was supported by NASA LST grant NAG5-3525 and NSF grant
AST-9802568 at OSU.
JRP gratefully acknowledges a Forchheimer Visiting
Professorship at HU.
Support to AVK was provided by NASA through Hubble Fellowship grant
HF-01121.01-99A from the Space Telescope Science Institute, which is
operated by the Association of Universities for Research in Astronomy,
Inc., under NASA contract NAS5-26555.


\def\re{\reference}

\vfill\eject

\bigskip\bigskip
\cl{\footnotesize
\begin{tabular}{ccccccccc}
\multicolumn{9}{l}{ {\bf Table 1:} Parameter fits for the mass
functions    } \\
\hline\hline
 & & \multicolumn{3}{c}{distinct halos} & &
\multicolumn{3}{c}{subhalos} \\
$z$ & & $\tal$ & $\log \phi_*^a$ & $\log M_*^b$ &  & $\tal$ &
$\log \phi_*^a$ & $\log M_*^b$ \\
0.00  & & $-0.85\pm 0.04$ & $-5.37\pm 0.41$ & $15.82\pm 16.24$ &
        & $-0.72\pm 0.10$ & $-5.35\pm 0.41$ & $15.03\pm 16.72$ \\
0.25 & & $-0.86\pm 0.04$ &  $-4.90\pm 0.34$ & $15.23\pm 16.26$ &
        & $-0.80\pm 0.10$ & $-5.52\pm 0.31$ & $14.95\pm 16.53$ \\
0.70 & & $-0.93\pm 0.06$  & $-5.24\pm 0.36$ & $15.37\pm 15.96$ &
        & $-0.89\pm 0.16$ & $-5.51\pm 0.41$ & $14.64\pm 15.56$ \\
1.35 & & $-1.00\pm 0.05$  & $-5.72\pm 0.32$ & $15.53\pm 15.82$ &
        & $-0.92\pm 0.15$ & $-5.90\pm 0.56$ & $14.80\pm 16.70$ \\
1.99 & & $-1.06\pm 0.05$  & $-5.91\pm 0.26$ & $15.46\pm 15.85$ &
        & $-0.85\pm 0.11$ & $-5.80\pm 0.32$ & $14.79\pm 16.78$ \\
2.95 & & $-1.16\pm 0.06$  & $-5.77\pm 0.33$ & $14.92\pm 15.81$ &
        & $-0.94\pm 0.18$ & $-6.08\pm 0.33$ & $14.54\pm 17.11$ \\
4.00 & & $-1.32\pm 0.09$  & $-6.26\pm 0.23$ & $14.68\pm 14.20$ &
        & $-0.65\pm 0.08$ & $-5.81\pm 0.42$ & $14.88\pm 16.85$ \\
4.92 & & $-1.29\pm 0.09$  & $-6.35\pm 0.28$ & $14.61\pm 14.85$ &
        & $-0.46\pm 0.15$ & $-5.72\pm 0.46$ & $15.05\pm 18.51$ \\
\hline
\hline
\multicolumn{9}{l}{$^a$ units of $[\hmpc ]^{-3}$ } \\
\multicolumn{9}{l}{$^b$ units of $\hmsun$ } \\
\label{tab:masstab}
\end{tabular}
}

\bigskip\bigskip

\cl{\footnotesize
\begin{tabular}{ccccc}
 \multicolumn{5}{l}{ {\bf Table 2:} Conditional mass function fits } \\
\hline
\hline
 & & \multicolumn{3}{c}{$M_{\rm host}^b < 10^{13}$} \\
$z$ & &  $\tal$ & $\log \phi_*^a$ & $\log M_*^b$  \\
0.00 & & $-0.71\pm 0.07$ & $-5.52\pm 0.37$ & $14.92\pm 16.08$ \\
0.08 & & $-0.80\pm 0.09$ & $-5.57\pm 0.28$ & $14.67\pm 16.08$ \\
0.25 & & $-0.72\pm 0.07$ & $-5.52\pm 0.39$ & $14.88\pm 16.35$ \\
0.41 & & $-1.06\pm 0.12$ & $-6.15\pm 0.27$ & $14.35\pm 15.80$ \\
0.50 & & $-0.95\pm 0.11$ & $-6.07\pm 0.35$ & $14.63\pm 16.07$ \\
0.70 & & $-0.87\pm 0.09$ & $-5.92\pm 0.47$ & $14.81\pm  15.96$ \\
1.35 & & $-0.87\pm 0.07$ & $-5.92\pm 0.26$ & $14.82\pm  15.97$ \\
1.99 & & $-0.66\pm 0.09$ & $-5.35\pm 0.39$ & $14.86\pm  15.71$ \\
2.95 & & $-0.98\pm 0.16$ & $-6.03\pm 0.51$ & $14.36\pm  17.39$ \\
\hline
 & & \multicolumn{3}{c}{$ 10^{13}<M_{\rm host}^b<10^{14}$} \\
$z$ & &  $\tal$ & $\log \phi_*^a$ & $\log M_*^b$  \\
0.00  & & $-0.61\pm 0.08$ & $-5.45\pm 0.21$ & $14.99\pm 16.70$ \\
0.08  & & $-0.68\pm 0.09$ & $-5.70\pm 0.55$ & $14.92\pm 16.90$ \\
0.25  & & $-0.91\pm 0.16$ & $-5.90\pm 0.52$ & $14.50\pm 16.58$ \\
0.41  & & $-0.69\pm 0.08$ & $-5.47\pm 0.31$ & $15.03 \pm 16.69$ \\
0.50  & & $-0.80\pm 0.15$ & $-6.03\pm 0.48$ & $14.76 \pm 16.69$ \\
0.70  & & $-0.58\pm 0.07$ & $-5.54\pm 0.40$ & $15.42 \pm 17.18$ \\
1.35  & & $-0.42\pm 0.10$ & $-6.00\pm 0.55$ & $16.58 \pm 17.32$ \\
\hline
 & & \multicolumn{3}{c}{$M_{\rm host}^b>10^{14}$} \\
$z$ & &  $\tal$ & $\log \phi_*^a$ & $\log M_*^b$  \\
0.00  & & $-0.65\pm 0.11$ & $-5.49\pm 0.55$ & $14.87 \pm  17.00$ \\
0.08  & & $-0.72\pm 0.09$ & $-5.80\pm 0.53$ & $14.96 \pm  16.57$ \\
0.25  & & $-0.64\pm 0.12$ & $-5.46\pm 0.63$ & $15.03 \pm  16.91$ \\
0.41  & & $-0.71\pm 0.11$ & $-5.94\pm 0.52$ & $14.90\pm  17.13$ \\
0.50  & & $-0.59\pm 0.17$ & $-5.57\pm 0.71$ & $14.93\pm 18.13$ \\
0.70  & & $-0.70\pm 0.10$ & $-5.86\pm 0.45$ & $14.84 \pm  16.32$ \\
1.35  & & $-0.78\pm 0.10$ & $-5.96\pm 0.34$ & $14.74 \pm 16.38$ \\
\hline
\hline
\multicolumn{5}{l}{$^a$ units of $[\hmpc ]^{-3}$ } \\
\multicolumn{5}{l}{$^b$ units of $\hmsun$ } \\
\label{tab:cond}
\end{tabular}
}

\bigskip\bigskip

\cl{
\begin{tabular}{cccccc}
\multicolumn{6}{l}{ {\bf Table 3:} Parameter fits for the velocity
functions  } \\
\hline
\hline
 & & \multicolumn{2}{c}{distinct halos} & \multicolumn{2}{c}{subhalos} \\
$z$ & & $\log \psi_*^a$ & $\tbe$ &  $\log \psi_*^a$ & $\tbe$ \\
0.00  & & $4.93\pm 0.06$ & $-2.82\pm  0.03$ & $4.22\pm 0.20$ &
$-3.04\pm 0.10$ \\
0.50  & & $6.25\pm 0.08$ & $-3.39\pm 0.04$ & $6.96\pm 0.39$
&$-4.16\pm 0.20$ \\
1.35  & & $5.86\pm 0.09$ & $-3.19\pm 0.04$ & $4.40\pm 0.38$ &
$-3.00\pm 0.18$ \\
2.31  & & $6.66\pm 0.11$ & $-3.51\pm 0.05$ & $4.03\pm 0.27$ &
$-2.84\pm 0.13$ \\
2.95  & & $6.84\pm 0.07$ & $-3.57\pm 0.03$ & $3.02\pm 0.21$ &
$-2.66\pm 0.10$ \\
4.00  & & $7.54\pm 0.10$ & $-3.96\pm 0.05$ & $0.76\pm 0.16$ &
$-1.62\pm 0.08$ \\
4.92  & & $6.00\pm 0.26$ & $-3.33\pm 0.12$ & & \\
\hline\hline
\multicolumn{6}{l}{$^a$ units of $[\hmpc ]^{-3}$ } \\
\label{tab:vel}
\end{tabular}
}

\end{document}